\documentclass[aps,pre,groupedaddress,preprint]{revtex4-1}
\usepackage{physics}
\usepackage{amsmath}
\usepackage{graphicx}
\usepackage{subfigure}
\usepackage{hyperref}
\usepackage{bm}
\begin{document}
\title{Drift instabilities driven by slab ion temperature gradient in suprathermal plasmas} 
\author{Ran Guo}
\thanks{Author to whom correspondence should be addressed}
\email{rguo@cauc.edu.cn}
\affiliation{Department of Physics, College of Science, Civil Aviation University of China, Tianjin 300300, China}
\begin{abstract}
The drift instabilities driven by the slab ion temperature gradient (ITG) in Kappa-distributed plasmas are investigated by the kinetic method.
The linear dispersion relation is given in an integral representation involving only the standard plasma dispersion function.
The wave frequency and growth rate are derived without the density inhomogeneity.
Numerical solutions of the dispersion equation are conducted to show the different effects of the suprathermal ions and electrons.
We find that the suprathermal ions can enhance the instability in large wavenumbers but suppress it in small wavenumbers. 
Thus, the suprathermalization of ions could be one of the factors leading to a lower limit of wavenumbers for the ITG instabilities.
Besides, the numerical calculations also imply that the thermal speed ratio affects the intensities of the suprathermal effects.
Finally, in the presence of density inhomogeneity, the ITG instability boundary is numerically analyzed.
\end{abstract}
\maketitle

\section{Introduction}
\label{sec:intro}
The ion-temperature-gradient (ITG) mode is a significant drift instability in magnetized plasmas and has been extensively studied in past decades. 
In fusion plasmas, the ITG mode is responsible for the anomalous transport of heat and particles \cite{Horton1999}. 
Coppi \textit{et al.} derived the growth rate of ITG instability in slab geometry \cite{Coppi1967}. 
Guo and Romanelli gave the threshold of the ITG mode in toroidal geometry \cite{Guo1993}. 
Gultekin and Gurcan solved the stable and unstable roots of the dispersion relation of toroidal ITG mode, and their results showed that the unstable branch is asymmetric near the instability threshold \cite{Gueltekin2018}. 
Ryter \textit{et al.} proved that the ion heat transport is enhanced when the normalized temperature gradient exceeds a critical value \cite{Ryter2019}. 
Wang \textit{et al.} studied the effect of the trapped fast ions on ITG mode and found that those ions can stabilize the ITG mode if the temperature exceeds a threshold \cite{Wang2022b}. 
In space plasmas, the temperature inhomogeneity is ubiquitous, and the ITG mode may exist.
Lu \textit{et al.} found that the ITG could be an intrinsic property in the Earth's magnetotail \cite{Lu2017}. 
Artemyev \textit{et al.} studied the profiles of ion density and temperature in the magnetotail and found that plasma temperature changes along and across the magnetotail \cite{Artemyev2017}.

Non-thermal particles are usually observed in space plasmas.
Kappa distribution is a widely used model to describe the suprathermal particles in various plasma environments, such as solar winds \cite{Maksimovic1997a,LynnB.Wilson2019}, the planetary magnetosphere \cite{Hapgood2011,Eyelade2021}, and the plasma sheet \cite{Christon1988,Christon1989}.
These suprathermal particles can significantly affect the behaviors of the waves and instabilities \cite{Guo2022a,Guo2022,Guo2023a,Lazar2023,Verheest2024}.

The Kappa-distributed plasmas and the ion temperature gradient can exist simultaneously in some plasma environments.
Espinoza \textit{et al.} reported that the Kappa-distributed electrons and ions are observed with inhomogeneous temperature in the magnetotail \cite{Espinoza2018}.
Christon \textit{et al.} found that the suprathermal electrons and ions can be fitted by the Kappa distribution very well in the plasma sheet \cite{Christon1988,Christon1989}.
Furthermore, the simulation \cite{Lu2017} revealed that the ion temperature gradient is nearly constant in the magnetotail.
These works indicate that the slab ITG mode in suprathermal plasmas is a possible physical process in the Earth's magnetotail.

The theory of ITG mode has been extended by considering the non-thermal particles.
Batool \textit{et al.} investigated the ITG mode in the presence of Kappa- and Cairns-distributed electrons and found that these non-thermal electrons strengthen the instabilities \cite{Batool2012}. 
Mirza \textit{et al.} derived the nonlinear equations for toroidal ITG modes with Kappa-distributed electrons and concluded that the non-thermal electrons can significantly modify the nonlinear vortex structure \cite{Mirza2015}. 
Naeem \textit{et al.} studied the ITG-driven vortices in $(r,q)$-distributed plasmas and showed that these non-thermal particles significantly affect the vortex size \cite{Naeem2020}.  
Zhou \textit{et al.} investigated the solitons driven by the ITG in regularized Kappa-distributed plasmas, and their results pointed out that the suprathermal electrons reduce the soliton amplitude but increase its width \cite{Zhou2022}.

Most previous studies investigated the ITG mode in the presence of non-thermal electrons, while the effects of suprathermal ions are rarely discussed.
However, the observations \cite{Collier1996,Chotoo2000,Espinoza2018,Eyelade2021} showed that the ions also follow the Kappa distribution in space plasmas.
Many investigations \cite{Guo2021a,Lazar2023,Guo2023a} implied that the suprathermalizations of different species play very different roles in affecting the plasma waves and instabilities.
Among these works, we studied the different suprathermal effects of electrons and ions on the drift modes driven by the density gradient \cite{Guo2023a}. 
It was found that the suprathermalization of electrons can cause a lower limit of wavenumbers for drift instability, which does not exist in Maxwellian plasmas.
The suprathermalization of ions only affects this critical wavenumber but does not determine its existence.

Therefore, the present work focus on the suprathermal effects of ions and electrons on the ITG mode by the kinetic method.
To this aim, the paper is organized as follows.
In Sec. \ref{sec:model}, we introduce the model of the ITG mode in Kappa-distributed plasmas and derive the general dispersion relation in slab geometry.
In Sec. \ref{sec:itg}, we study the ITG instability without the density gradient and analyze the effects of the suprathermal ions and electrons.
In Sec. \ref{sec:boundary}, we investigate the influence of the suprathermal particles on the boundary of the ITG instability in the presence of density inhomogeneity.
Finally, the conclusions are given in Sec. \ref{sec:sum}.

\section{Model and Linear Dispersion Relations in Slab Geometry}
\label{sec:model}
We consider a collisionless plasma composed of Kappa-distributed ions and electrons in a uniform magnetic field $\vb{B}_0=B_0\vb{e}_z$.
The ion number density and temperature are assumed to be inhomogeneous along the $x$-axis, and their gradients are supposed to be weak and constant.
We assume the quasi-neutrality condition $n_e \approx n_i$ holds, implying that the electron number density is also non-uniform.
However, the electron temperature is considered uniform because the ITG modes propagate much slower than the electron thermal speed.
The stationary distribution function is supposed to be \cite{Guo2023a}, 
\begin{equation}
    f_\sigma^{(0)} = \frac{n_\sigma(X_\sigma)}{[2\pi \kappa_\sigma \theta_\sigma^2(X_\sigma)]^{\frac{3}{2}}}
    \frac{\Gamma(\kappa_\sigma+1)}{\Gamma(\kappa_\sigma-1/2)} 
    \left[1+\frac{v^2}{2 \kappa_\sigma \theta^2_\sigma(X_\sigma)}\right]^{-\kappa_\sigma-1},
    \label{eq:f0}
\end{equation}
where $n_\sigma$ is the number density of species $\sigma$, $\theta_\sigma$ is the thermal speed, $\kappa_\sigma$ is the kappa parameter, and $X_\sigma=x+y/\Omega_\sigma$ is the motion constant of the Lamor cyclotron with the gyrofrequency $\Omega_\sigma=q_\sigma B/m_\sigma$.
The temperature can be derived from the second moment of the distribution function, 
\begin{equation}
    T_\sigma = \frac{1}{3}\int m_\sigma v^2 \frac{f^{(0)}_\sigma}{n_\sigma} \dd{\vb{v}} = \frac{\kappa_\sigma}{\kappa_\sigma-3/2} m_\sigma \theta_\sigma^2.
    \label{eq:T}
\end{equation}
It should be mentioned that the distribution for a real plasma is more complex than the Kappa distribution \eqref{eq:f0} assumed here.
Espinoza \textit{et al.} fitted the observation data by the Kappa distribution with temperature variations in the magnetotail \cite{Espinoza2018}.
Their analyses showed that both the temperatures and the kappa parameters for electrons and ions change with spatial variations during different physical processes.
Therefore, the distribution considered in this work could be regarded as a simplified model of the real plasmas, focusing on the suprathermal effects on the ITG instabilities.

In such a plasma model, the linear dispersion relation of electrostatic modes could be derived under the weak inhomogeneity assumption and the local approximation \cite{Basu2008,Guo2023a,Mikhailovskii1974}. 
In the previous work \cite{Guo2023a}, we derived a novel integral representation of the drift dispersion relation involving only the standard dispersion function defined in Maxwellian plasmas. 
If one considers the linear mode propagating in the $y$-$z$ plane with the low-frequency condition $\omega \ll \Omega_i$, the dispersion relation reads \cite{Guo2023a}, 
\begin{multline}
    \varepsilon = 1+\sum_\sigma \frac{1}{k^2 \lambda_{\kappa\sigma}^2} \Bigg\{ 1+
    \int_0^{+\infty} \dd{b_\sigma} 
    \left[G_1(b_\sigma) - G_2(b_\sigma) \frac{\hat{\omega}_{d\sigma}}{\omega}\right] \\
    \times
    \left[
    W\left(\frac{\omega}{k_z\theta_\sigma/\sqrt{b_\sigma}}\right) -1
    \right]
    \Lambda_0 \left(\frac{k_y^2 \rho_\sigma^2}{b_\sigma}\right)
    \Bigg\}=0,
    \label{eq:dr-kappa-w-lf}
\end{multline}
where $\lambda_{\kappa\sigma}$ is the Debye length in suprathermal plasmas \cite{Lazar2022a}, 
\begin{equation}
   \lambda_{\kappa\sigma} = \sqrt{\frac{\kappa_\sigma}{\kappa_\sigma-1/2}}
   \sqrt{\frac{\epsilon_0 m_\sigma \theta_\sigma^2}{n_\sigma q_\sigma^2}},
   \label{eq:l_D}
\end{equation}
$W(z)$ is the plasma dispersion $W$-function \cite{Weiland2000},  $\Lambda_0(x)=\exp(-x)I_n(x)$ is the product of the exponential function with the modified Bessel function $I_n(x)$, 
and $\rho_\sigma = \theta_\sigma/|\Omega_\sigma|$ is the average Larmor radius for $\sigma$-particles.
$G_1(b_\sigma)$ and $G_2(b_\sigma)$ are two weighed functions,
\begin{equation}
    G_1(b_\sigma) = \frac{\kappa_\sigma^{\kappa_\sigma+1/2}b_\sigma^{\kappa_\sigma-1/2}}{\Gamma(\kappa_\sigma+1/2)} e^{-\kappa_\sigma b_\sigma},
    \label{eq:G1}
\end{equation}
\begin{equation}
    G_2(b_\sigma) = \left(\kappa_\sigma-\frac{3}{2}\right)\frac{\kappa_\sigma^{\kappa_\sigma-1/2}b_\sigma^{\kappa_\sigma-3/2}}{\Gamma(\kappa_\sigma+1/2)} e^{-\kappa_\sigma b_\sigma}.
    \label{eq:G2}
\end{equation}
The drift frequency operator is defined as,
\begin{equation}
    \hat{\omega}_{d\sigma} = \frac{k_y T_\sigma}{\Omega_\sigma m_\sigma} \left( \left. \dv{\ln n_\sigma}{x} \right|_ {x=0} + \left. \dv{T_\sigma}{x} \right|_{x=0} \pdv{}{T_\sigma} \right).
    \label{eq:wd}
\end{equation}
In the limit $\kappa_\sigma \rightarrow \infty$, the Debye length (\ref{eq:l_D}) reduces to the standard Debye length $\lambda_{D\sigma}$ in Maxwellian plasmas, and the weighted functions $G_1(b_\sigma)$ and $G_2(b_\sigma)$ reduce to the delta function $\delta(b_\sigma-1)$ (see Appendix A in Ref. \cite{Guo2023a}), restoring the dispersion relation (\ref{eq:dr-kappa-w-lf}) to the standard drift dispersion relation in Maxwellian plasmas \cite{Guo2023a}. 

After inserting the drift frequency operator (\ref{eq:wd}) into the dispersion relation (\ref{eq:dr-kappa-w-lf}) and calculating the partial derivative $\pdv*{}{T_\sigma}$, one obtains the dielectric function,
\begin{equation}
    \varepsilon = 1 +\frac{1}{k^2\lambda_{\kappa e}^2}(1+K_e)+\frac{1}{k^2 \lambda_{\kappa i}^2} (1+K_{i1}+K_{i2}),
    \label{eq:dr-itg}
\end{equation}
where
\begin{equation}
    K_e = \int_0^{+\infty} \dd{b_e} \left[G_1(b_e)-G_2(b_e)\frac{\omega_{Ne}}{\omega}\right]\left[ W(\xi_e\sqrt{b_e}) -1 \right] \Lambda_0 \left(\frac{k_y^2 \rho_e^2}{b_e}\right),
    \label{eq:Ke}
\end{equation}
\begin{equation}
    K_{i1} = \int_0^{+\infty} \dd{b_i} \left[ G_1(b_i)- G_2(b_i) \frac{\omega_{Ni}-\omega_{Ti}/2}{\omega} \right] \left[ W(\xi_i\sqrt{b_i}) -1 \right] \Lambda_0 \left(\frac{k_y^2 \rho_i^2}{b_i}\right),
    \label{eq:Ki1}
\end{equation}
\begin{multline}
    K_{i2} = -\int_0^{+\infty} \dd{b_i} G_2(b_i) \frac{\omega_{Ti}}{\omega}
    \Bigg\{  
    \frac{\xi_i^2}{2} b_i W(\xi_i\sqrt{b_i}) \Lambda_0\left(\frac{k_y^2 \rho_i^2}{b_i}\right) \\
    + \left[W(\xi_i\sqrt{b_i}) -1 \right]
    \left[ 
        \Lambda_1 \left(\frac{k_y^2 \rho_i^2}{b_i}\right) - \Lambda_0 \left(\frac{k_y^2 \rho_i^2}{b_i}\right) 
    \right]
    \frac{k_y^2 \rho_i^2}{b_i}
    \Bigg\},
    \label{eq:Ki2}
\end{multline}
with the abbreviations $\xi_{e,i}=\omega/(k_z\theta_{e,i})$, and the drift frequencies,
\begin{equation}
    \omega_{Ni} = \frac{k_y T_i}{\Omega_i m_i} \frac{1}{L_N} = \frac{\kappa_i}{\kappa_i-3/2}\frac{k_y \theta^2_i}{\Omega_i} \frac{1}{L_N},
    \label{eq:wni}
\end{equation}
\begin{equation}
    \omega_{Ne} = \frac{k_y T_e}{\Omega_e m_e} \frac{1}{L_N} = \frac{\kappa_e}{\kappa_e-3/2}\frac{k_y \theta^2_e}{\Omega_e} \frac{1}{L_N},
    \label{eq:wne}
\end{equation}
\begin{equation}
    \omega_{Ti} = \frac{k_y T_i}{\Omega_i m_i} \frac{1}{L_T} = \frac{\kappa_i}{\kappa_i-3/2}\frac{k_y \theta^2_i}{\Omega_i} \frac{1}{L_T},
    \label{eq:wti}
\end{equation}
where $L_N^{-1}=(\dv*{\ln n_e}{x})_{x=0}=(\dv*{\ln n_i}{x})_{x=0}$ and $L_T^{-1}=(\dv*{\ln T_i}{x})_{x=0}$ are the the characteristic lengths of inhomogeneous density and temperature, respectively.

It should be noticed that some previous works \cite{Beskin1987,Fo.1989} showed that the dispersion relation derived by expanding the distribution function up to the first order of inhomogeneity may contain undesirable features.
Beskin \textit{et al} \cite{Beskin1987} suggested an effective permittivity describing correctly the energy exchange between waves and particles. 
For the drift modes with the $x$-direction inhomogeneity and $z$-direction uniform magnetic field, the effective dielectric function can be obtained by \cite{Beskin1987} $\varepsilon^{\text{eff}} = \varepsilon + (i/2) \pdv*{\varepsilon}{k_x}{x}$,
where $\varepsilon$ is the standard dielectric function derived in the present work. 
However, this study focuses on the wave propagating in the $y$-$z$ plane, so the second term in the effective dielectric function vanishes due to the zero partial derivative with respect to $k_x$.
Therefore, our result \eqref{eq:dr-itg} is still valid.

\section{ITG instability in the absence of density inhomogeneity}
\label{sec:itg}
\subsection{Wave frequencies and growth rates}
In the case of $\nabla n_i = \nabla n_e = 0$, the dispersion relation \eqref{eq:dr-itg}-\eqref{eq:Ki2} can be analytically solved by neglecting the ion gyroradius $\rho_i$.
Because the ITG mode propagates within the speed range $\theta_i \ll \omega_R/k_z \ll \theta_e$, 
one can use the following approximations \cite{Ichimaru2004}, 
\begin{equation}
    W(\xi_e\sqrt{b_e}) \approx 1, \quad \text{for }\xi_e \ll 1,
    \label{eq:W_approx_e}
\end{equation}
and
\begin{equation}
    W(\xi_i\sqrt{b_i}) \approx -\frac{1}{\xi_i^2 b_i} -\frac{3}{\xi_i^4 b_i^2}, \quad \text{for }\xi_i \gg 1.
    \label{eq:W_approx_i}
\end{equation}
So, the dispersion equation $\varepsilon=0$ turns into,
\begin{equation}
    1 + \frac{k_z^2 c_s^2}{\omega^2} \left(\frac{\omega_{Ti}}{\omega}-1\right)=0,
    \label{eq:epsilon_0}
\end{equation}
where $c_s$ is the ion sound speed in Kappa-distributed plasmas \cite{Mace1998}, 
\begin{equation}
    c_s=\sqrt{\frac{\kappa_e}{\kappa_e-1/2}}\sqrt{\frac{m_e}{m_i}}\theta_e.
    \label{eq:cs}
\end{equation}
For the low-frequency condition $\omega \ll \omega_{Ti}$, the complex frequency is solved from Eq. \eqref{eq:epsilon_0},
\begin{equation}
    \omega = \left(\frac{1}{2}+i\frac{\sqrt{3}}{2}\right)\left(\omega_{Ti}k_z^2 c_s^2\right)^{1/3}.
    \label{eq:wc}
\end{equation}
Although the complex frequency \eqref{eq:wc} is of the same form as that in Maxwellian plasmas \cite{Mikhailovskii1974}, 
the ion sound speed $c_s$ \eqref{eq:cs} is enhanced by the suprathermal electrons, and the drift frequency $\omega_{Ti}$ \eqref{eq:wti} is amplified by the increased temperature due to the suprathermal ions.
Therefore, both Kappa-distributed electrons and ions can enhance the ITG instability.

To accurately illustrate the effects of the suprathermal particles on the ITG instability, we numerically calculate the dispersion equation $\varepsilon=0$ with Eqs. \eqref{eq:dr-itg}-\eqref{eq:wti}.
For convenience, dimensionless parameters are adopted. 
The real frequency and growth rate are scaled by the ion cyclotron frequency $\Omega_i$,
while the wavenumbers are normalized by the ion Larmor radius $\rho_i$.
The choice of the kappa parameters is based on the observations.
The kappa parameter of electrons was observed to be $2<\kappa_e<5$ in the solar wind \cite{Maksimovic1997a}, $4\le \kappa_e \le 5$ in the magnetosphere \cite{Eyelade2021}, and $3<\kappa_e<6$ in the plasma sheet \cite{Espinoza2018}. 
For ions, the kappa value was reported to be $2.4<\kappa_i<4.7$ in the solar wind \cite{Collier1996,Chotoo2000} and $4<\kappa_i<10$ in the plasma sheet \cite{Espinoza2018}. 
In addition, Eyelade \textit{et al.} found $5 \le \kappa_i \le 8$ in the magnetosphere for most cases, but the minimum value of $\kappa_i$ could be 1.51 \cite{Eyelade2021}.
Therefore, the kappa values of electrons and ions are investigated in a typical range of $1.8<\kappa_{e,i}<5.5$ in this work. 
The mass ratio is set as $m_i/m_e=1836$.
The ratio of thermal speed is studied in the range $100 \le \theta_e/\theta_i \le 200$.
We set the ion plasma frequency much larger than the ion cyclotron frequency $\omega_{pi}/\Omega_i=100$,
which is motivated by the observed values \cite{Yoon2018b}.
The characteristic length of inhomogeneous temperature is assumed to be $L_T/\rho_i=10$ if not specified.

\begin{figure}
    \centering
    \includegraphics[width=0.7\textwidth]{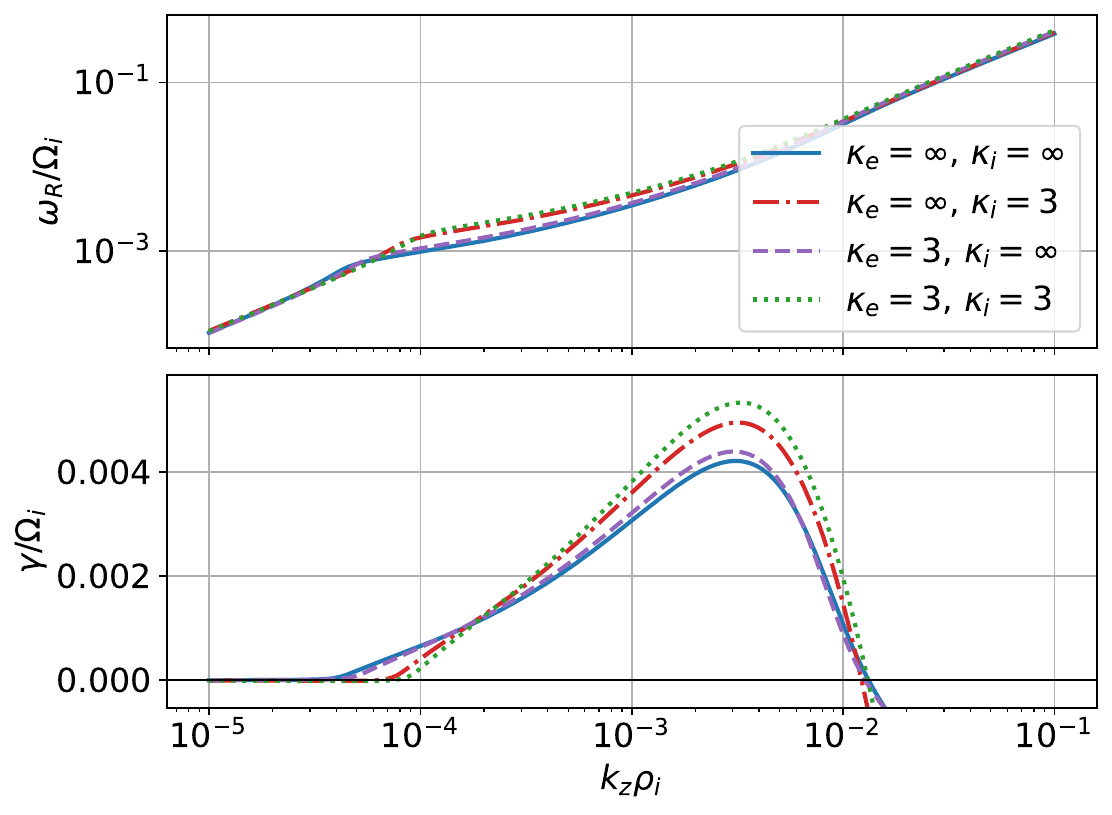}
    \caption{The wave frequency (upper panel) and growth rate (lower panel) of the ITG mode.
    The $y$-direction wavenumber $k_y\rho_i = 0.1$ and the thermal speed ratio $\theta_e/\theta_i = 150$ are set.
    }
    \label{fig1:dr-itg}
\end{figure}

Figure \ref{fig1:dr-itg} illustrates the real frequency and growth rate of the ITG mode as functions of the wavenumber $k_z\rho_i$ for four sets of kappa parameters: (a) $\kappa_e=\kappa_i=\infty$, (b) $\kappa_e=\infty$ and $\kappa_i=3$, (c) $\kappa_e=3$ and $\kappa_i=\infty$, and (d) $\kappa_e=\kappa_i=3$.
The thermal speed ratio $\theta_e/\theta_i=150$ is assumed.
The suprathermal effect of ions can be observed by comparing the cases (a) (blue solid lines) and (b) (red dotted-dashed lines). 
The real frequency is increased by the suprathermal ions in the wavenumbers roughly $10^{-4}<k_z\rho_i<10^{-2}$ but nearly unchanged in the other wavenumbers.
The instability (the positive growth rate) is enhanced by the suprathermalization of ions in large wavenumbers but suppressed in small wavenumbers.
The effects of suprathermal electrons can be found by comparing cases (a) (blue solid lines) and (c) (purple dashed lines).
As shown in both the upper and lower panels, one can conclude that the suprathermalization of electrons only slightly affects the wave frequency and growth rate, indicating that the suprathermal electrons are less important than the suprathermal ions in the ITG mode.
The total suprathermal effect, plotted by the green dotted lines, can be understood as a summation of the impacts of suprathermal electrons and ions.

The effects of suprathermal ions exhibited in Fig. \ref{fig1:dr-itg} can be explained as follows.
As is known, the slab ITG instability is a deformation of the ion-acoustic wave due to the temperature gradient \cite{Mikhailovskii1974}. 
Therefore, the growth rate is determined by the competition between the ion Landau damping and the destabilization effects from the ion diamagnetic drift driven by the temperature gradient.
On the one hand,
the suprathermalization of ions raises drift frequency \eqref{eq:wti} and thus strengthens ion diamagnetic drift, intensifying the ITG instability.
To be more specific, by substituting Eqs. \eqref{eq:wti} and \eqref{eq:cs} into Eq. \eqref{eq:wc}, the theoretical growth rate can be rewritten in the unit of the ion cyclotron frequency,
\begin{equation}
    \frac{\gamma}{\Omega_i} = \frac{\sqrt{3}}{2} \left(\frac{\kappa_i}{\kappa_i-3/2}\frac{\kappa_e}{\kappa_e-1/2} k_y\rho_i k_z^2\rho_i^2 \frac{\rho_i}{L_T} \frac{m_e\theta_e^2}{m_i\theta_i^2}\right)^{1/3},
    \label{eq:gamma2wci}
\end{equation}
indicating that the ITG instability could be enhanced due to a reduced $\kappa_i$ for a non-zero $k_z\rho_i$.
On the other hand,
when $k_z\rho_i$ approaches zero, the suprathermal ions increase the instability only to a small extent but largely amplify the Landau damping \cite{Mace1998}, reducing the ITG instability in small wavenumbers.

\begin{figure}
    \centering
    \includegraphics[width=0.7\textwidth]{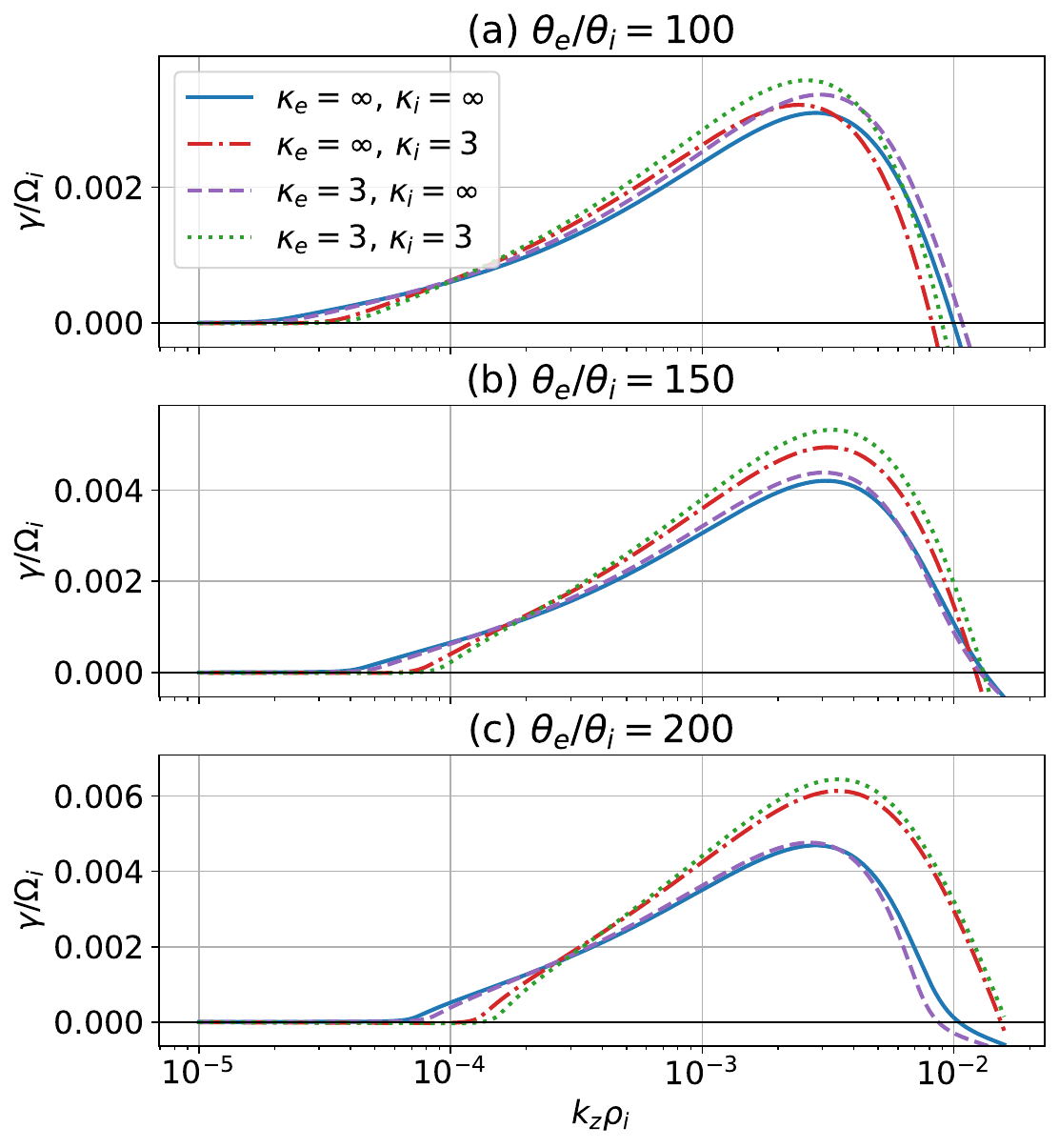}
    \caption{The ITG growth rate vs $k_z\rho_i$ for different thermal speed ratios (a) $\theta_e/\theta_i=100$, (b) $\theta_e/\theta_i=150$, and (c) $\theta_e/\theta_i=200$.
    The $y$-direction wavenumber is $k_y\rho_i = 0.1$.
    }
    \label{fig2:gamma-dif-theta}
\end{figure}

The suprathermal effects are also influenced by the thermal speed ratio.
Figure \ref{fig2:gamma-dif-theta} illustrates the growth rate of the ITG mode as a function of the wavenumber $k_z\rho_i$ for three different thermal speed ratios $\theta_e/\theta_i=100$, $150$, and $200$.
It shows that the suprathermal ions have a more noticeable effect on the ITG instability when the thermal speed ratio is higher, but the suprathermal electrons take the opposite effect.
It could be explained by the dispersion equation \eqref{eq:dr-itg}.
As the thermal speed ratio $\theta_e/\theta_i$ increases, the ion contribution to the dielectric function becomes much larger, while the electron one becomes much smaller, which is attributed to the modified Debye length \eqref{eq:l_D}. 
Therefore, the ion suprathermalization becomes more significant when the electron-to-ion thermal speed ratio is much higher.
When the thermal speed ratio decreases to a relatively small value,
the magnitudes of the suprathermal effects from ions and electrons are comparable,
as shown in the upper panel of Fig. \ref{fig2:gamma-dif-theta}. 

\subsection{The wavenumber range of the instabilities}
\label{sec:kz-range}

\begin{figure}
    \centering
    \includegraphics[width=\textwidth]{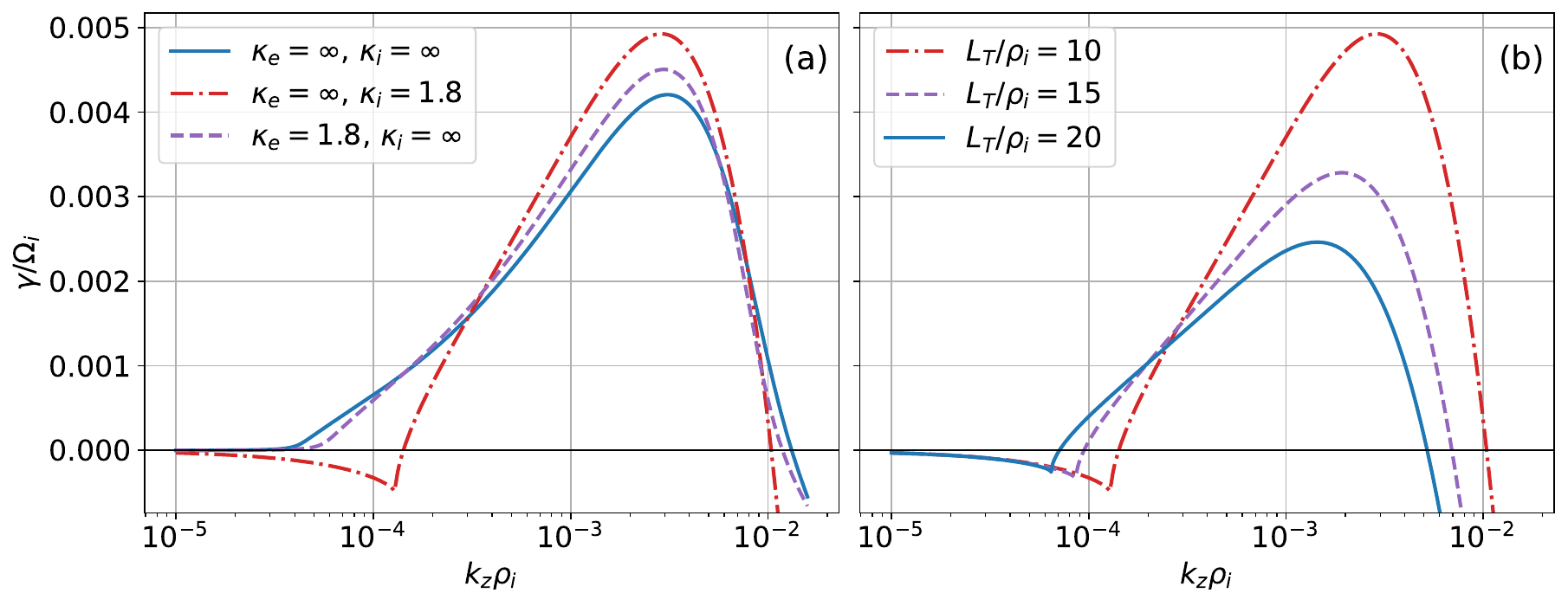}
    \caption{The ITG growth rate for (a) different kappa parameters and (b) different characteristic lengths of temperature inhomogeneity.
    In the two subplots, the $y$-direction wavenumber $k_y\rho_i = 0.1$ and the thermal speed ratio $\theta_e/\theta_i = 150$ are set.
    In subfigure (a), the characteristic length of inhomogeneous temperature is $L_T/\rho_i=10$.
    In subfigure (b), the kappa parameters are $\kappa_e = \infty$ and $\kappa_i = 1.8$.
    It is worth noting that the red dotted-dashed lines in the two subfigures correspond to the same parameters.
    }
    \label{fig3:dr-small-kappa}
\end{figure}

As explained in the previous subsection, the ITG instability is a combination of the destabilization from the ion diamagnetic drift and the stabilization from the ion Landau damping.
Therefore, if the suprathermalization of ions is sufficiently strong, i.e. a sufficiently small $\kappa_i$, the ITG mode might be stable in small $z$-direction wavenumbers due to a strong enough Landau damping.
This case is illustrated in Fig. \ref{fig3:dr-small-kappa}(a).
It shows that the strong suprathermalization of ions leads to an extra lower limit of wavenumbers for the ITG instabilities, but the suprathermalization of electrons does not. 
However, ion suprathermalization is only one of the factors determining the existence of such a lower limit of wavenumbers.
As is shown in the subsequent Fig. \ref{fig5:dr-kykz}, the propagating direction of the wave is another factor.

Although the suprathermalization of ions can stabilize the ITG mode in small wavenumbers, the instability does not vanish.
The reason is as follows.
On the one hand, 
a sufficient small $\kappa_i$ decreases the growth rate to a negative value in small wavenumbers but enhances the growth rate in large wavenumbers,
as shown in Fig. \ref{fig3:dr-small-kappa}(a).
On the other hand, if the kappa parameters are fixed,
the maximum growth rate can be reduced by two factors due to Eq. \eqref{eq:gamma2wci}, i.e., the decline in the thermal speed ratio $\theta_e/\theta_i$ and the increment of the characteristic length of inhomogeneous temperature $L_T/\rho_i$.
The decline in $\theta_e/\theta_i$ results in a negligible suprathermal effect, as shown in Fig. \ref{fig2:gamma-dif-theta}, and the ITG instability in slab geometry has no threshold in Maxwellian plasmas \cite{Mikhailovskii1974}.
The increment of $L_T/\rho_i$ reduces the maximum growth rate while it also diminishes the absolute value of the negative growth rate in small wavenumbers,
as shown in Fig. \ref{fig3:dr-small-kappa}(b).
Therefore, the instability still survives even with strong suprathermalization of ions and a non-zero ion temperature gradient.
The suprathermal effects just change the wavenumber range for the ITG instabilities but cannot give rise to a threshold.

\begin{figure}
    \centering
    \includegraphics[width=0.7\textwidth]{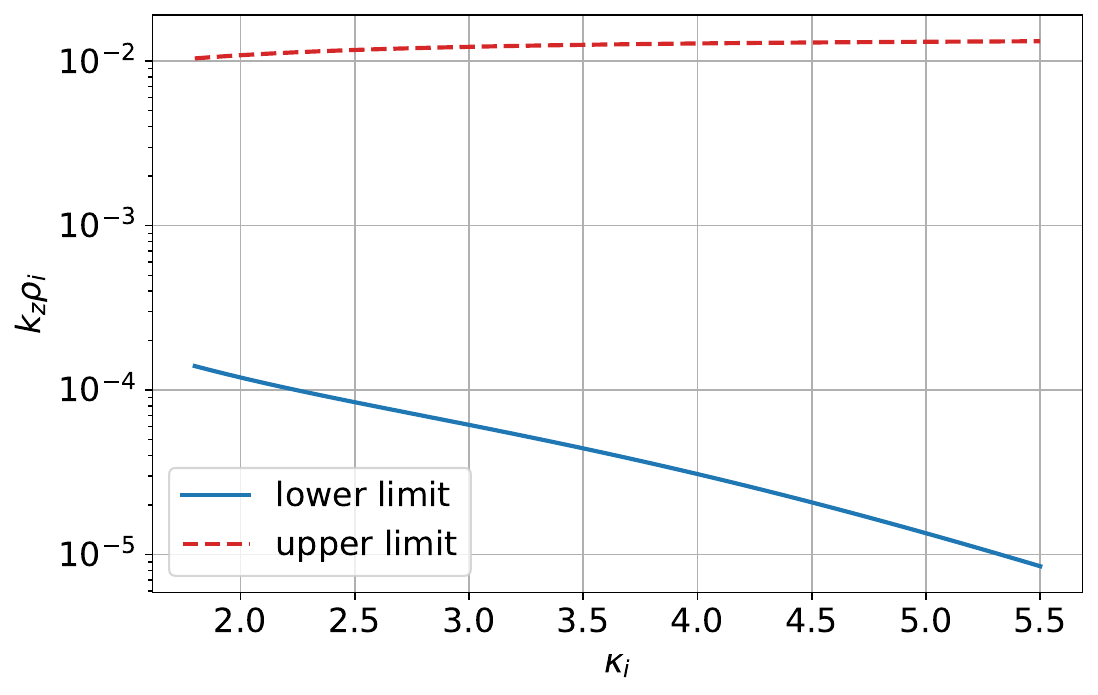}
    \caption{The lower and upper limit of wavenumbers for the ITG instability. The Maxwellian electrons are chosen ($\kappa_e=\infty$).
    The $y$-direction wavenumber $k_y\rho_i = 0.1$ and the thermal speed ratio $\theta_e/\theta_i = 150$ are adopted.
    }
    \label{fig4:kz-cr}
\end{figure}

The critical wavenumber $k_{z \mathrm{lim}} \rho_i$ for $\gamma=0$ could be solved from the dispersion equation $\varepsilon=0$ with Eqs. \eqref{eq:dr-itg}-\eqref{eq:Ki2} (see Appendix \ref{ap:kz-cr} for details).
We plot the lower and the upper limit of the wavenumbers as functions of $\kappa_i$ for $k_y\rho_i=0.1$ and $\theta_e/\theta_i=150$ in Fig. \ref{fig4:kz-cr}.
In this figure, we only consider the case of Kappa-distributed ions and Maxwellian electrons because it is the suprathermal ions determining the existence of the lower limit in wavenumbers.
It shows that the lower limit of wavenumbers decreases with an increasing $\kappa_i$, but the upper limit is nearly unchanged.
In addition, one may wonder why such a lower limit of wavenumbers does not appear in Figs. \ref{fig1:dr-itg} and \ref{fig2:gamma-dif-theta}.
The reason is that the $\gamma/\Omega_i$ is negative but very close to zero in those figures, which is hardly observed.

\begin{figure}
    \centering
    \includegraphics[width=0.7\textwidth]{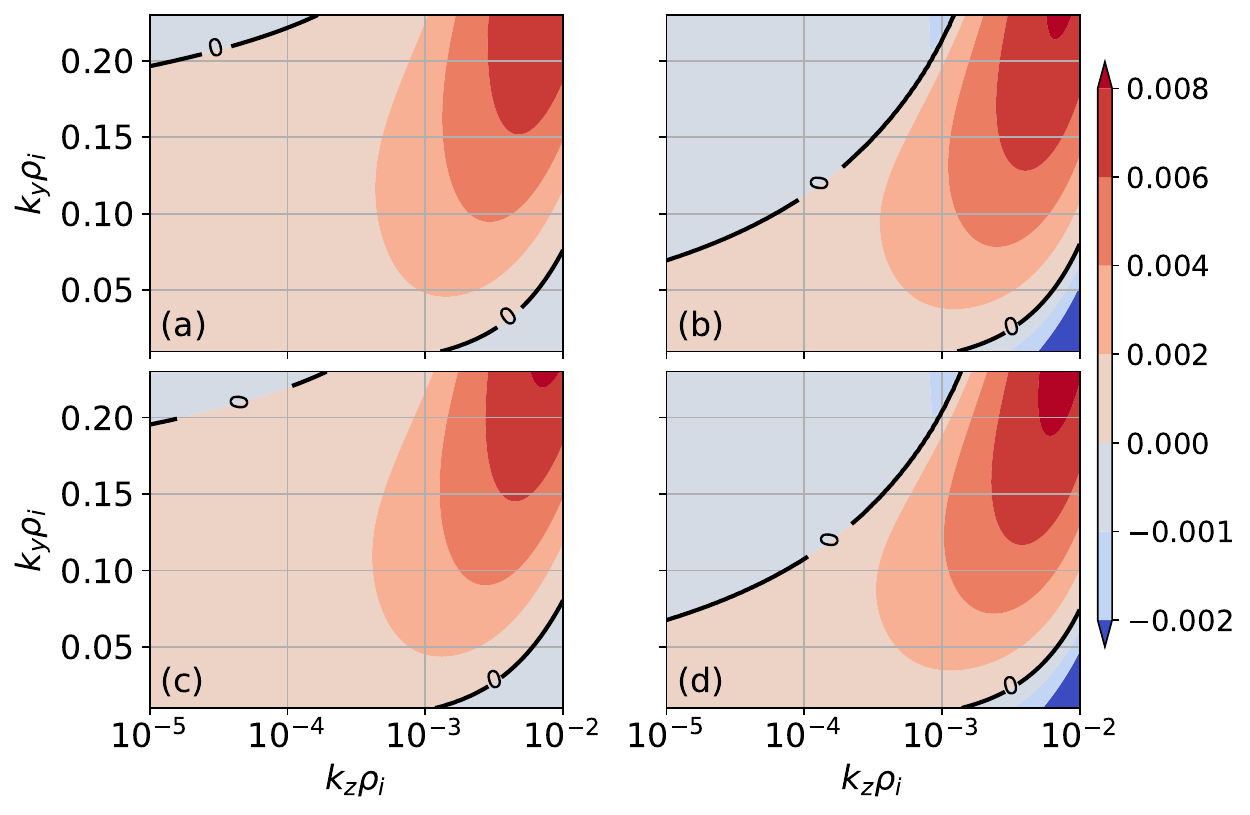}
    \caption{The ITG growth rate vs varied $k_y\rho_i$ and $k_z\rho_i$ for (a) $\kappa_e=\kappa_i=\infty$, (b) $\kappa_e=\infty$, $\kappa_i=3$, (c) $\kappa_e=3$, $\kappa_i=\infty$, and (d) $\kappa_e=\kappa_i=3$.
    The thermal speed ratio is $\theta_e/\theta_i = 150$.
    }
    \label{fig5:dr-kykz}
\end{figure}

Figure \ref{fig5:dr-kykz} plots the growth rate for varied wavenumbers in $y$ and $z$ directions.
It reflects that the boundary of the stable regime is altered by the suprathermal ions but hardly changed by the suprathermal electrons.
Therefore, the suprathermalization of ions is a determined factor affecting the wavenumber range of ITG instabilities.
Due to the ion suprathermalization, the ITG instability exists in a more narrow range of wavenumbers in Kappa-distributed plasmas than in thermal plasmas.
It is worth noting that the lower limit of wavenumbers is determined not only by the ion suprathermalization but also by the propagating direction of the wave.
Even if the plasmas are Maxwellian, the ITG mode is unstable only in a certain range of $k_y$ and $k_z$, as shown in Fig. \ref{fig5:dr-kykz}(a).

\section{The ITG instability boundary in the presence of density inhomogeneity}
\label{sec:boundary}

\begin{figure}
    \centering
    \includegraphics[width=0.7\textwidth]{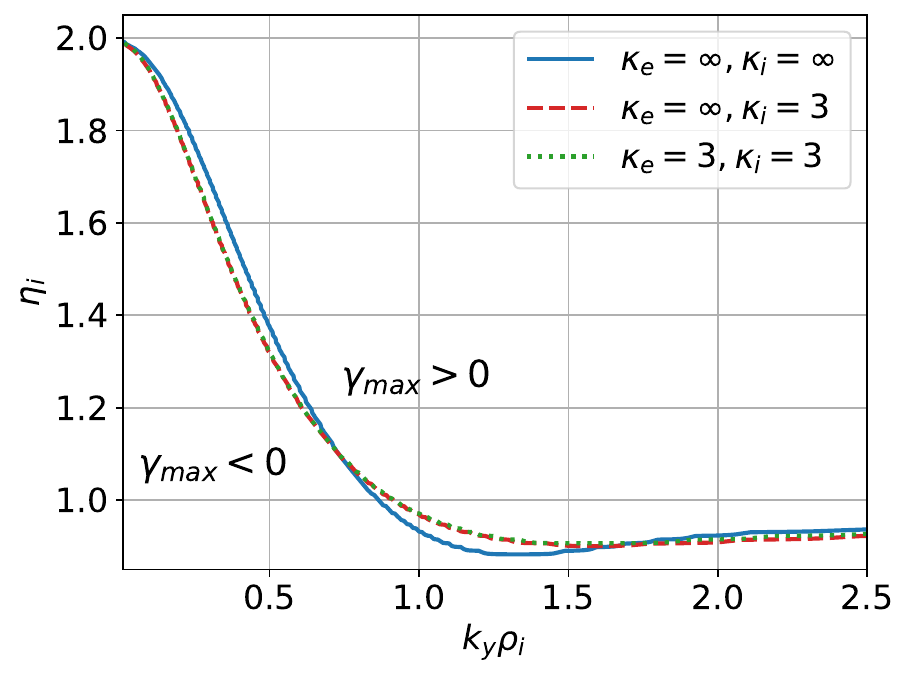}
    \caption{The critical value of $\eta_i$ for the ITG instability.
    The maximum growth rate $\gamma_{max}$ is solved in the $z$-direction wavenumber range $10^{-4}<k_z\rho_i<10^{-2}$.
    The thermal speed ratio is $\theta_e/\theta_i=150$.
    }
    \label{fig6:boundary}
\end{figure}

If the ion density gradient is involved, the ITG instability could arise for a large enough ratio of the ion temperature gradient to the density gradient $\eta_i = (\dd{\ln T_i}/\dd{x})/(\dd{\ln n_i}/\dd{x})$ \cite{Mikhailovskii1974,Horton1999}. 
The reason is that the ion would stabilize the wave if only the density gradient exists in Maxwellian and Kappa-distributed plasmas \cite{Mikhailovskii1974,Guo2023a}. 
Therefore, there is an instability boundary between $\eta_i=0$ and $\eta_i=\infty$.
The numerical results of the critical value of $\eta_i$ for the ITG instability are shown in Fig. \ref{fig6:boundary}.
We find that the suprathermal ions just affect the critical value of $\eta_i$ to a small extent, and the suprathermal electrons hardly change the boundary.
It is attributed to the fact that the suprathermal ions only change the wavenumbers of the instability but do not eliminate the instabilities, as shown in Figs. \ref{fig1:dr-itg}-\ref{fig5:dr-kykz}.

Besides, we must stress that the growth rates calculated in Fig. \ref{fig6:boundary} are only for the ITG modes.
When $\eta_i \rightarrow 0$, the ion temperature gradient vanishes, but the density gradients of ions and electrons remain.
In this limiting case, the ITG mode becomes stable, while the density drift mode is unstable due to the inhomogeneous electron density.
The ITG modes and the density drift modes are two different branches of the drift waves.
However, the latter is not calculated in Fig. \ref{fig6:boundary} because the former is the main concern of this study.

\section{Summary}
\label{sec:sum}
In this paper, we study the ITG instability in Kappa-distributed plasmas.
The linear dispersion relation is given by Eqs. \eqref{eq:dr-itg}-\eqref{eq:wti}.
In the absence of density inhomogeneity, the analytical wave frequency and growth rate are derived in Eq. \eqref{eq:wc}, respectively.
To illustrate the suprathermal effects more clearly, we numerically calculate the dispersion equation and plot the wave frequency and growth rate in Figs. \ref{fig1:dr-itg}-\ref{fig5:dr-kykz}.
It reveals that the suprathermal effects of ions and electrons are very different.
First, the suprathermal ions enhance the ITG instability in large wavenumbers but suppress it in small wavenumbers, while the suprathermal electrons only slightly affect the instability.
Second, the thermal speed ratio $\theta_e/\theta_i$ influences the strengths of the suprathermal effects.
Third, the strong suprathermalization of ions is one of the factors leading to a lower limit of wavenumbers for the instabilities.
Another factor determining the lower limit of wavenumbers is the wave propagating direction.
Finally, the suprathermal effects on the boundary of the ITG instability are also investigated in the presence of density inhomogeneity.

As a final remark, both the simulation \cite{Lu2017} and the observation \cite{Espinoza2018} showed the existence of the slab ion temperature gradient in the Earth's magnetotail.
Moreover, the electrons and ions are observed to follow the Kappa distribution in this area \cite{Espinoza2018,Christon1988,Christon1989}.
Therefore, the present work may be applied to the studies of plasma behaviors in such space environments.

\begin{acknowledgments}
This work was supported by the National Natural Science Foundation of China (No. 12105361) and by the Scientific Research Startup Foundation of Civil Aviation University of China (No. 2015QD05X).
\end{acknowledgments}

\section*{Data Availability}
No new data were created or analyzed in this study.

\appendix
\section{Derivation of the critical \texorpdfstring{$k_z\rho_i$}{kz ρi} for \texorpdfstring{$\gamma=0$}{γ=0} in the case of Kappa-distributed ions and Maxwellian electrons}
\label{ap:kz-cr}
Due to the reason explained in the Sec. \ref{sec:kz-range}, we consider the case of Kappa-distributed ions and Maxwellian electrons. 
To find the critical wavenumber $k_{z \mathrm{lim}} \rho_i$, we need to solve the dispersion equation $\varepsilon=0$ with Eqs. \eqref{eq:dr-itg}-\eqref{eq:Ki2} for $\gamma=0$.
In this case, the complex frequency becomes real, $\omega=\omega_R$, 
and the dielectric function can be divided into the real and imaginary parts with the aid of \cite{Ichimaru2004}, 
\begin{equation}
    W(z) = 1 - z e^{-z^2/2} \int_0^z e^{t^2/2} \dd{t} + i \sqrt{\frac{\pi}{2}} z e^{-z^2/2}.
    \label{eq:W-re-im}
\end{equation}
Substituting Eq. \eqref{eq:W-re-im} into Eqs. \eqref{eq:dr-itg}-\eqref{eq:Ki2} and using the approximations \eqref{eq:W_approx_e} and $k\lambda_{D e} \ll 1$, we derive the real part of the dielectric function,
\begin{equation}
    \Re \varepsilon = \frac{1}{k^2\lambda_{D e}^2} + \frac{1}{k^2 \lambda_{\kappa i}^2} 
    \left[ 1 + A_1(\xi_i) + \frac{\omega_{Ti}}{2\omega_R}A_2(\xi_i) + \frac{\omega_{Ti}}{\omega_R}A_3(\xi_i) \right],
    \label{eq:re-e}
\end{equation}
with the definitions,
\begin{align}
    A_1(\xi_i) =& \int_0^{+\infty} \dd{b_i} G_1(b_i) \left[\Re W(\xi_i\sqrt{b_i}) - 1 \right] \Lambda_0 \left(\frac{k_y^2 \rho_i^2}{b_i}\right),\label{eq:A1} \\ 
    A_2(\xi_i) =& \int_0^{+\infty} \dd{b_i} G_2(b_i) \left[\Re W(\xi_i\sqrt{b_i}) - 1 \right] \Lambda_0 \left(\frac{k_y^2 \rho_i^2}{b_i}\right), \label{eq:A2} \\ 
    A_3(\xi_i) =& \int_0^{+\infty} \dd{b_i} G_2(b_i) \Big\{
    -\frac{\xi_i^2}{2} b_i \Re W(\xi_i\sqrt{b_i}) \Lambda_0\left(\frac{k_y^2 \rho_i^2}{b_i}\right) \\ \notag
    &+ \left[\Re W(\xi_i\sqrt{b_i}) -1 \right] \left[\Lambda_0\left(\frac{k_y^2 \rho_i^2}{b_i}\right) - \Lambda_1\left(\frac{k_y^2 \rho_i^2}{b_i}\right)\right] \frac{k_y^2 \rho_i^2}{b_i}
    \Big\}.  \label{eq:A3}
\end{align}
The imaginary part of the dielectric function \eqref{eq:dr-itg} can be written as,
\begin{equation}
    \Im \varepsilon = \sqrt{\frac{\pi}{2}}\frac{\xi_e}{k^2\lambda_{D e}^2} + \sqrt{\frac{\pi}{2}}\frac{\xi_i}{k^2 \lambda_{\kappa i}^2} 
    \left\{
        B_1(\xi_i) + \frac{\omega_{Ti}}{\omega_R}\left[\frac{B_2(\xi_i)}{2} - \frac{\xi_i^2}{2}B_3(\xi_i) - k_y^2 \rho_i^2B_4(\xi_i) \right]
    \right\},
    \label{eq:im-e}
\end{equation}
where $\exp(-\xi_e^2/2) \approx 1$ is used for $\xi_e \ll 1$, and the coefficients are defined as,
\begin{align}
    & B_1(\xi_i) = \int_0^{+\infty} \dd{b_i} 
    G_1(b_i) b_i^{1/2} \exp\left(-\frac{\xi_i^2 b_i}{2}\right)
    \Lambda_0 \left(\frac{k_y^2 \rho_i^2}{b_i}\right),\label{eq:B1} \\
    & B_2(\xi_i) = \int_0^{+\infty} \dd{b_i} 
    G_2(b_i) b_i^{1/2} \exp\left(-\frac{\xi_i^2 b_i}{2}\right)
    \Lambda_0 \left(\frac{k_y^2 \rho_i^2}{b_i}\right),\label{eq:B2} \\
    & B_3(\xi_i) = \int_0^{+\infty} \dd{b_i} 
    G_2(b_i) b_i^{3/2} \exp\left(-\frac{\xi_i^2 b_i}{2}\right)
    \Lambda_0 \left(\frac{k_y^2 \rho_i^2}{b_i}\right),\label{eq:B3} \\
    & B_4(\xi_i) = \int_0^{+\infty} \dd{b_i} 
    G_2(b_i) b_i^{-1/2} \exp\left(-\frac{\xi_i^2 b_i}{2}\right)
    \left[ \Lambda_1 \left(\frac{k_y^2 \rho_i^2}{b_i}\right) - 
    \Lambda_0 \left(\frac{k_y^2 \rho_i^2}{b_i}\right) \right]. \label{eq:B4}
\end{align}
Collecting $\Re \varepsilon = 0$ and $\Im \varepsilon = 0$, one solves the critical wavenumber $k_{z \mathrm{lim}} \rho_i$,
\begin{equation}
    k_{z \mathrm{lim}} \rho_i = - \frac{\omega_{Ti}}{\Omega_i} \frac{1}{\xi_{\mathrm{lim}}} 
                                \frac{A_2(\xi_{\mathrm{lim}})/2+A_3(\xi_{\mathrm{lim}})}{\frac{\kappa_i}{\kappa_i-1/2}\frac{m_i\theta_i^2}{m_e\theta_e^2}+A_1(\xi_{\mathrm{lim}})+1},
    \label{eq:kzlim}
\end{equation}
where $\xi_{\mathrm{lim}} = \omega_R/(k_{z\mathrm{lim}}\theta_i)$ is the ratio between wave speed and thermal speed at $\gamma=0$.
The $\xi_{\mathrm{lim}}$ is implicitly determined by the following equation (which is also obtained from $\Re \varepsilon = 0$ and $\Im \varepsilon = 0$),
\begin{multline}
    \frac{\kappa_i}{\kappa_i-1/2} \frac{m_i\theta_i^3}{m_e\theta_e^3} + B_1(\xi_{\mathrm{lim}})
    -\frac{\frac{\kappa_i}{\kappa_i-1/2}\frac{m_i\theta_i^2}{m_e\theta_e^2}+A_1(\xi_{\mathrm{lim}})+1}{A_2(\xi_{\mathrm{lim}})/2+A_3(\xi_{\mathrm{lim}})} \\
    \times \left[\frac{B_2(\xi_{\mathrm{lim}})}{2} - B_3(\xi_{\mathrm{lim}})\frac{\xi_{\mathrm{lim}}^2}{2} - B_4(\xi_{\mathrm{lim}}) k_y^2 \rho_i^2\right]=0.
    \label{eq:xi-eq}
\end{multline}

\bibliography{library}
\end{document}